\documentclass[letterpaper, preprint, paper, 11pt]{AAS}	

\usepackage[utf8]{inputenc}

\usepackage{graphicx}
\usepackage{amsmath}
\usepackage[mathscr]{euscript}
\usepackage[version=4]{mhchem}
\usepackage{siunitx}
\usepackage{longtable,tabularx}
\usepackage{float}
\usepackage{placeins}

\usepackage[utf8]{inputenc}
\usepackage{amssymb}
\usepackage{placeins}
\usepackage{bm}
\usepackage{siunitx}
\usepackage{graphicx}
\usepackage{placeins}
\usepackage{graphicx}
\usepackage{float}
\usepackage{footnpag}		
\usepackage{overcite}
\usepackage{bm}
\usepackage{amsmath}
\usepackage{subcaption}
\usepackage{caption}
\usepackage[linesnumbered,ruled,vlined]{algorithm2e}
\usepackage{algpseudocode}
\usepackage{comment}
\usepackage{tikz}
\usepackage{footnpag}			      	
\usepackage{comment}
\usepackage{xcolor}
\usepackage{amsthm}

\SetKwInput{KwInput}{Input}                
\SetKwInput{KwOutput}{Output}              

\newcommand{\reals}{{\mathbb R}}
\newcommand{\sphere}{{\mathbb S}}
\newcommand{\SOThree}{\textrm{SO}(3)}
\newcommand{\soThree}{\mathfrak{so}(3)}

\newcommand{\nnreals}{\mathbb{R}_{\geq 0}}
\newcommand{\stateSpace}{\mathcal{X}}
\newcommand{\inputSpace}{\mathcal{U}}

\newcommand{\deputyState}{{\mathbf x}}

\newcommand{\deputyStateMat}{{\mathbf A}}
\newcommand{\meanmotion}{\eta}
\newcommand{\Azimuth}{\text{Az}}
\newcommand{\Elevation}{\text{El}}

\newcommand{\chiefStatex}{\mathbf{x}_0}

\newcommand{\chiefInertia}{{\mathcal J}}
\newcommand{\chiefInertiaInv}{[{\mathcal J}]^{-1}}
\newcommand{\chiefInput}{u}
\newcommand{\traj}{\xi}
\newcommand{\dynfunc}{f}
\newcommand{\constraintfunc}{h}
\newcommand{\controlfunc}{c}
\newcommand{\initial}{\theta}

\newcommand{\deputyUncertainty}{\mathbf{P}}
\newcommand{\noiseInput}{\mathbf{G}}
\newcommand{\processNoise}{\mathbf{Q}}
\newcommand{\obsMat}{C}
\newcommand{\obsState}{\mathbf{y}}
\newcommand{\pointing}{\mathbf{p}}

\newcommand{\aov}{\alpha}

\newcommand{\skewMat}[1]{\mathcal{S}({#1})}
\newcommand{\idMat}[1]{\mathbf{I}_{#1 \times #1}}
\newcommand{\zeroMat}[2]{\mathbf{0}_{#1 \times #2}}
\newcommand{\emptyMat}[2]{\mathbf{\emptyset}_{#1 \times #2}}
\newcommand{\trace}{\mathrm{trace}}

\newcommand{\rotmat}[2]{\mathcal{R}_{#1 \to #2}}
\newcommand{\dotrotmat}[2]{\dot{\mathcal{R}}_{#1\to#2}}
\newcommand{\hillsframe}{\mathcal{H}}
\newcommand{\bodyframe}{\mathcal{B}}
\newcommand{\inertialframe}{\mathcal{I}}
\newcommand{\restr}{\downarrow}

\newcommand{\A}{\mathcal{A}}
\newcommand{\B}{\mathcal{B}}
\newcommand{\C}{\mathcal{C}}

\newcommand{\cost}{\mathcal{J}}

\newcommand{\switchtime}{\tau}
\newcommand{\sunvector}{\mathbf{s}}

\newcommand{\roll}{\phi}
\newcommand{\pitch}{\theta}
\newcommand{\yaw}{\psi}
\newcommand{\angularvel}[3]{\boldsymbol{\omega}_{#1#2}^{#3}}

\newcommand{\dotangularvel}[3]{\dot{\boldsymbol{\omega}}_{#1#2}^{#3}}

\newcommand{\norm}[1]{|#1|}

\PaperNumber{23-P12} 

\title{Autonomous Local Catalog Maintenance of Close Proximity Satellite Systems on Closed Natural Motion Trajectories\thanks{Approved for public release; distribution is unlimited. Public Affairs approval AFRL20230485}}
\author{Christopher W. Hays\thanks{Ph.D. Candidate, Aerospace Engineering Department, Embry-Riddle Aeronautical University, 1 Aerospace Blvd., Daytona Beach, FL, 32114.}, 
\  Kristina Miller\thanks{Ph.D. Candidate, Electrical and Computer Engineering Department, University of Illinois Urbana-Champaign, 1308 W Main St, Urbana, IL 61801},
\ Alexander Soderlund\thanks{Research Aerospace Engineer, Space Vehicles Directorate, Air Force Research Laboratory, 3550 Aberdeen Dr. SE, Kirtland AFB, NM}, 
\ Sean Phillips \thanks{Research Mechanical Engineer, Space Vehicles Directorate, Air Force Research Laboratory, 3550 Aberdeen Dr. SE, Kirtland AFB, NM}
\ Troy Henderson \thanks{Associate Professor, Aerospace Engineering Department, Embry-Riddle Aeronautical University, 1 Aerospace Blvd., Daytona Beach, FL., 32114.}
}

\begin{document}
\maketitle

\thispagestyle{plain}
\pagestyle{plain}

\begin{abstract}
     To enable space mission sets like on-orbit servicing and manufacturing, agents in close proximity maybe operating too close to yield resolved localization solutions to operators from ground sensors. 
     This leads to a requirement on the systems need to maintain a catalog of their local neighborhood, however, this may impose a large burden on each agent by requiring updating and maintenance of this catalog at each node. 
     To alleviate this burden, this paper considers the case of a single satellite agent (a chief) updating a single catalog. 
     More specifically, we consider the case of numerous satellite deputy agents in a local neighborhood of a chief, the goal of the chief satellite is to maintain and update a catalog of all agents within this neighborhood through onboard measurements. 
     We consider the agents having relative translational and attitude motion dynamics between the chief and deputy, with the chief centered at the origin of the frame. 
     We provide an end-to-end solution of the this problem through providing both a supervisory control method coupled with a Bayesian Filter that propagates the belief state and provides the catalog solutions to the supervisor. 
     The goal of the supervisory controller is to determine which agent to look at and at which times while adhering to constraints of the chief satellite. 
     We provide a numerical validation to this problem with three agents. 
\end{abstract}






\section{Introduction}
Space is a key component to the global economy by facilitating trillions of dollars, annually, in the global market due to GPS alone \cite{GPS_FinalReport}. Due to the lower barrier to entry, space is becoming increasingly congested and the current space surveillance network may not be able to fully maintain the entire catalog of small objects in the future \cite{SGOBBA2013411}. 
Therefore, future satellites may have to maintain a local catalog of nearby space objects within close proximity to assure safety of the systems and autonomously decide if any evasive actions would be required. 
Recently, readiness levels for space trusted autonomy are studied.\cite{hobbsSTARL2023}
In this paper, we consider the problem of local area catalog maintenance for satellite systems. 
This work may be particularly useful for such mission types as in space servicing and manufacturing, wherein a ``foreman satellite'' is tasked with maintaining local awareness of a formation of satellite robots. 

In the literature, the topic of satellite localization has been thoroughly studied for close proximity (relative motion) satellite systems \cite{opromolla2017pose,patel2012relative,5259193,geller2014angles, hu2020three, doi:10.2514/2.4379,5530989}, and ground-based sensor networks \cite{schutz2004statistical, sciotti2011low}. 
Due to the number of space objects compared to the number of ground based sensors, the tasking of ground sensors must be intelligently chosen \cite{herz2017ssa,jaunzemis2016evidence}. 
Situational awareness, in general, can be defined as ``the perception of the elements in the environment within a volume of time and space, the (organizational) comprehension of their meaning, and the projection of their status in the near future" \cite{endsley2017toward}. 
In particular, \textit{space} situational awareness is a growing topic of interest \cite{holzinger2018challenges}. 
Space situational awareness (a.k.a. space domain awareness, or SDA) has been expanded to encompass all space environmental impacts as well, which leads to a multidisciplinary domain of research that incorporates facets of information fusion, collection-tasking and exploitation to better assess anomaly  identification and prediction. 

In this article, we consider the case of a single controllable chief satellite equipped with an electro-optical sensor and on-board estimating capabilities that must track and maintain the states of several uncontrollable deputy satellites, each of which lay on a closed, elliptical natural motion trajectory \cite{frey2017constrained} about the chief. The local catalog we wish to maintain is the concatenated list of all the deputies' positional and velocity states relative to the chief. However, a deputy's state is inherently uncertain. Generally, these state uncertainties
are represented through stochastic parameters, such as a mean and state covariance matrix pair for a Gaussian probability distribution, also known as the ``belief state." Through the use of a Bayesian filter \cite{crassidis2004optimal}, the fusion between a deputy's prior belief state and an observation of that deputy results in a posterior belief state with reduced uncertainty. Crucially, the chief cannot observe the states of all deputies simultaneously and is only able to make measurements of deputies when they lie within the chief's field-of-view. Thus, the uncertainties of unobserved agents will grow accordingly with the amount of time elapsed without measurement. This engenders the core of the local catalog maintenance problem: how does the chief know when and where to look in order to provide a catalog of deputy states that are closest to their true values? 

Further, this article provides a decision-making algorithm which supplies a control trajectory to the chief satellite such that the uncertainties of all tracked deputy agents are guaranteed to lie underneath a predefined bound. 
The deputies are modeled to maneuver about the chief according to the well-known Clohessey-Wiltshire dynamics, while the chief's attitudinal dynamics are modeled according to gas jet thruster equations. 
The state estimation is performed using the extended Kalman filter, and the sequence of control inputs to the chief are computed via Model Predictive Control (MPC). 
Finally, a supervising decision logic framework is constructed that determines which sequence of local agents to observe based on the estimated relative positions and uncertainty. 
This decision logic framework is general enough to be paired alongside a graph-based approach to a multi-agent version of a similar problem presented by Hays et. al. \cite{hays2023enabling}.

The remainder of the paper continues with a section providing essential mathematical notation and definitions, an overview of the model predictive control process, and a description of the local catalog maintenance problem. The succeeding section details the proposed autonomous supervisor solution to the maintenance problem. The effectiveness of this method is then demonstrated through numerical simulations of the local maintenance scenario, followed by concluding remarks.



\section{Preliminaries}

\subsection{Notations and Definitions}
\label{sec:notation}
Let $\reals$ denote the set of all real numbers and let $\mathbb{N}$ denote the set of natural numbers. A frame for an $n$-dimensional space is defined by $n$ orthonormal vectors $\hat{\inertialframe}_i$, $i \in [1,n]$.
In this paper, we work with spaces where $n=3$. The Lie Group $\SOThree$ is the set of all real invertible $3 \times 3$ matrices that are orthogonal with determinant $1$, in this work referred to as \textit{rotation matrices} $\rotmat{}{} \in \SOThree:$
$$\SOThree := \{\rotmat{}{} \in \reals^{3 \times 3} \ | \ (\rotmat{}{})^{T}\rotmat{}{} = \idMat{3}, \ \det[\rotmat{}{}] = 1\}$$
Given two coordinate frames $\A$ and $\B$, the rotation of vector $\deputyState^{\A}$ as measured in frame $\A$ to $\deputyState^{\B}$ as measured in frame $\B$ is denoted by $\rotmat{\A}{\B}$,
that is, $\deputyState^{\B} = \rotmat{\A}{\B} \deputyState^{\A}$. Given three frames $\A$, $\B$, and $\C$, the angular velocity of $\B$ relative to $\A$ as measured in frame $\C$ is denoted by 
$\angularvel{\A}{\B}{\C}$. For a vector $\textbf{a} := (a_1 \ a_2 \ a_3)^{\top} \in \reals^3$, the cross product operator $\skewMat{\cdot} : \reals^{3} \to \soThree$:
\begin{equation}
\skewMat{\textbf{a}} =
\begin{pmatrix} 
0 & -a_3 & a_2 \\
a_3 & 0 & -a_1 \\
-a_2 & a_1 & 0
\end{pmatrix}
\label{eq:skewmat}
\end{equation}
If $\textbf{b} \in \reals^3$, then the cross-product $\textbf{c} = \textbf{a} \times \textbf{b}$ can be written $\textbf{c} = \skewMat{\textbf{a}}\textbf{b} = -\skewMat{\textbf{b}}\textbf{a}$. Here, the Lie algebra $\soThree$ is the set of all real $3 \times 3$ skew-symmetric matrices and is the tangent space of $\SOThree$ at the identity $\idMat{3}$:
$$\soThree := \{\skewMat{\textbf{a}}^{T} = -\skewMat{\textbf{a}} \ | \ \textbf{a} \in \reals^3\}$$
 For a vector $\textbf{a} \in \reals^3$, the $\mathcal{L}^2$ norm operator is defined as $\norm{\textbf{a}} := \sqrt{\textbf{a}^{\top} \textbf{a}}$. A multivariate Gaussian distribution of an $n$-dimensional random variable is parameterized by a mean vector $\hat{\deputyState} \in \reals^n$ and a symmetric, positive semidefinite covariance matrix $\deputyUncertainty \in \reals^{n \times n}$. Given matrices $A \in \reals^{m \times n}$ and $B \in \reals^{p \times q}$,
the Kroenecker product is denoted by $\otimes$ and is given by
$$
A \otimes B = \begin{pmatrix}
a_{11} B & a_{12} B & \dots & a_{1n} \\
a_{21} B & a_{22} B & \dots & a_{2n} \\
\vdots & \vdots & \ddots & \vdots \\
a_{m1} B & a_{m2} B & \dots & a_{mn} B 
\end{pmatrix}.
$$
Given a vector $a \in \reals^{m}$, the projection of $a$ onto $\reals^n$, $n \leq m$ is denoted by $a \restr \reals^n$. Next, we specify a mapping between the vector $\deputyState$ in Cartesian space to the  azimuth-elevation or (Az,El) space $g(\deputyState) : \reals^{3} \to \sphere^{2}$ 
\begin{equation}\label{eq:azel_angles}
    g(\deputyState) = 
    \begin{bmatrix}
    \Azimuth \\
    \Elevation
    \end{bmatrix} =
    \begin{bmatrix}
    \tan^{-1}(\frac{s_{2}}{s_{1}})\\
    \sin^{-1}(s_{3})
    \end{bmatrix}
\end{equation}
where $s=\frac{\deputyState}{\norm{\deputyState}}$ and $\sphere^2$ is the unit sphere with angular coordinates $\Azimuth \in [-\pi, \pi]$ and $\Elevation \in [-\frac{\pi}{2}, \frac{\pi}{2})$ by astronomical convention. 

\subsection{Model Predictive Control}
\label{sec:MPC}
Consider the following control system with state $x_k \in \reals^{n}$ at timestep $k$ that evolves according to a discrete-time system with process function $f : \reals^n \times \reals^m \to \reals^n$ and measurement function $c : \reals^n \times \reals^m \to \reals^{\ell}$ over timesteps $\{k, k+1, k+2, \dots\}$, $k \geq 0$ as
\begin{align}
\label{eq:discrete_system_dynamics}
x_{k+1} &= f(x_k, u_k) \\
\label{eq:discrete_system_measurement}
y_{k} &= c(x_k, u_k) 
\end{align}
where $u_k \in \reals^m$ is the input applied to the plant while $y_k \in \reals^{\ell}$, $\ell \leq n$, is the output of the system. Without loss of generality, we assume $f(0,0) = 0$ and $c(0,0) = 0$ for all $k \geq 0$. Let $N\in\mathbb{N}$, $N \geq 2$ be the finite horizon length. We define a control policy $\textbf{u}_{k} := \{u_{0}, u_{1}, \dots, u_{i}, \dots, u_{N-1}\}$ at time $k$ with an additional index $i = 0, \dots, N-1$. 
The output trajectory resulting from $\textbf{u}_{k}$ according to \eqref{eq:discrete_system_dynamics}-\eqref{eq:discrete_system_measurement} is $\textbf{y}_k := \{y_{0}, y_{1},\dots, y_{i}, \dots, y_{N-1}\}$. 
Let $\textbf{y}_k(i)$ and $\textbf{u}_k(i)$ denote particular elements of $\textbf{y}_k$ and $\textbf{u}_k$ at index $i$ given an initial condition $x_k$, respectively. 
Let there be some nonnegative function $h : \reals^{\ell} \times \reals^m \to \reals_{\geq 0}$, known as a stage cost, and a control objective function $J : \reals^{\ell} \times \reals^m \to \reals_{\geq 0}$ defined as
\begin{equation}
\label{eq:cost_function_generic}
J(\textbf{y}_k,\textbf{u}_k) := \sum^{N-1}_{i=0} h(\textbf{y}_k(i), \textbf{u}_k(i)).
\end{equation}
The goal of a model predictive controller is to, at timestep $k$, generate an \textit{optimal} control sequence $\textbf{u}^{*}_{k} := \{u^{*}_{0}, u^{*}_{1}, \dots, u^{*}_{i}, \dots, u^{*}_{N-1}\}$ such that the objective function $J$ from \eqref{eq:cost_function_generic} is minimized. The first input in this sequence, $\textbf{u}^{*}_{k}(0)$, is applied to the system. At timestep $k+1$, this minimization process is repeated and $\textbf{u}^{*}_{k+1}(0)$ is applied. This is formally written as
\begin{align}
\label{prob:MPC}
\begin{split}
\underset{\textbf{u}_k}{\textnormal{minimize}} & \ J(\textbf{y}_k,\textbf{u}_k)\\
\textnormal{subject to} &\  x_0 = x_k \\
                        &\  x_{i+1} = f(x_i,u_i), \quad i=0,\dots, N-1 \\
                        &\ y_{i} = c(x_i, u_i), \quad i=0,\dots, N-1 \\
                        &\ (x_{i}, u_{i}) \in \mathscr{X} \times \mathscr{U}, \quad i=0,\dots, N-1 \\
\end{split}
\end{align}
where $\mathscr{X} \subseteq \reals^n$ is the closed set of state constraints and $\mathscr{U}$ is the closed set of input constraints. Generally, the stage cost function $h$ is chosen to represent a distance of the output from some predefined desired output trajectory, i.e. $\textbf{y}^{*}_k$.

\section{Problem Statement}
\subsection{The Local Catalog Maintenance Problem}
Consider a chief spacecraft with state space $\stateSpace_0 \subseteq \reals^{n}$ and input space $\inputSpace \subseteq \reals^m$.
The state $\chiefStatex \in \stateSpace_0$
evolves according to $\dot{\mathbf{x}}_0 = \dynfunc(\chiefStatex, \chiefInput)$ where 
$\chiefInput \in \inputSpace$ is the chief's input,
and $\dynfunc$ is the chief's dynamic function.
Additionally, consider some constraint function $\constraintfunc : \stateSpace_0 \times \inputSpace \to \reals^{r}$.
A state $\chiefStatex \in \stateSpace_0$
and input $\chiefInput \in \inputSpace$
satisfies the constraints when $\constraintfunc_i(\chiefStatex, \inputSpace) \leq 0$
for each row $i \in \{1, \dots, r\}$ in the constraint vector.
Given an initial state $\initial_0 \in \stateSpace_0$
and an input function $\chiefInput : \nnreals \to \inputSpace$,
a {\em trajectory} of the chief is given by $\traj_{u,\initial_0} : \nnreals \to \stateSpace$ and satisfies 
$\traj_{u,\initial_0}(0) = \initial_0$ and
$\frac{d}{dt} \traj_{u, \initial_0}(t) = f(\traj_{u, \initial_0}(t), u(t))$.

The chief is tasked with maintaining a catalog of $d$ deputies.
Each deputy $i \in \{1, \dots, d\}$ has
a state space $\stateSpace_i \subseteq \reals^p$.
A deputy's state $\deputyState_i \in \stateSpace_i$ evolves according to $\dot{\deputyState}_i = \deputyStateMat \deputyState_i$
where $\deputyStateMat \in \reals^{p \times p}$ is the state matrix.
The state of all deputies, or system state, is given by $\deputyState := [\deputyState_i^{\top} \ \dots \ \deputyState_d^{\top}]^{\top}$
and evolves according to $\deputyState = (\idMat{d} \otimes \deputyStateMat) \deputyState$.

The chief is equipped with a recursive state estimator to track $d$ deputy belief states, parameterized by concatenated mean $\hat{\deputyState} := [\deputyState^{\top}_1 \ \cdots \ \deputyState^{\top}_d]^{\top}$ and covariance block matrix $\deputyUncertainty := \text{diag}\{\deputyUncertainty_1, \hdots,\deputyUncertainty_d\}$. 
These constructs evolve via their assumed dynamics models according to
\begin{gather}
    \dot{\hat{\deputyState}} = (\idMat{d} \otimes \deputyStateMat) \hat{\deputyState} \nonumber \\
    \dot{\deputyUncertainty} = (\idMat{d} \otimes \deputyStateMat) \deputyUncertainty (\idMat{d} \otimes \deputyStateMat)^{\top} + \noiseInput \processNoise \noiseInput^{\top},
    \label{eq:kalman-filter}
\end{gather}
where, $\processNoise \in \reals^{pd \times pd}$ is the process noise matrix and $\noiseInput \in \reals^{pd \times pd}$ is the noise input matrix. We assume $\noiseInput = \idMat{pd}$ for simplicity.

The chief is equipped with a sensor parameterized by its pointing vector in the chief's body frame $\pointing^{\bodyframe}$
and its angle of view $\aov$.
The sensor can observe the states of the deputies within some sensing field-of-view (FOV) dependent on
$\pointing^{\bodyframe}$, $\aov$,
and the chief's state $\chiefStatex$.
The linear measurement model of the sensor is given by
$\obsState = \obsMat(\deputyState, \chiefStatex) \deputyState$
where $\obsState$ is the observed state of deputies,
and $\obsMat$ is the observation matrix dependent on
the system state $\deputyState$
and the chief's state $\chiefStatex$.
The observation matrix $\obsMat$ is a block matrix given as
 \begin{equation}
    \obsMat(\deputyState, \chiefStatex) = \begin{bmatrix}
    \obsMat_1(\deputyState_1, \chiefStatex) & \emptyMat{p}{p} & \cdots & \emptyMat{p}{p}\\
    \emptyMat{p}{p} & \obsMat_2(\deputyState_2, \chiefStatex) & \cdots & \emptyMat{p}{p}\\
    \vdots & \vdots & \ddots & \vdots\\
    \emptyMat{p}{p} & \emptyMat{p}{p} & \cdots & \obsMat_d(\deputyState_d, \chiefStatex)
    \end{bmatrix},
    \label{eq:obsMat}
\end{equation}
where each $\obsMat_i(\deputyState_i, \chiefStatex)$ is
given as
\begin{equation}
    \obsMat_i(\deputyState_i, \chiefStatex) = \begin{cases}
        \idMat{p} & \textrm{if $\deputyState_i$ is observed} \\
        \emptyMat{p}{p} & \textrm{else.}
    \end{cases}
    \label{eq:deputy_obs}
\end{equation}
It should be noted here that $\emptyMat{p}{p}$ is an overloaded operator.
In the case that $\obsMat_i(\deputyState_i, \rotmat{\bodyframe}{\inertialframe}) = \idMat{p}$, $\emptyMat{p}{p}$ takes on the value of $\zeroMat{p}{p}$, and in the case when $\obsMat_i(\deputyState_i, \rotmat{\bodyframe}{\inertialframe}) = \emptyMat{p}{p}$, $\emptyMat{p}{p}$ is the null set taking the form of an empty matrix.

Whenever the chief makes an observation of a deputy $i \in \{1, \dots, d\}$
at a time instant $k$,
the measurement data is incorporated by the on-board estimator to impulsively update the belief state such that
$$(\hat{\deputyState}_i(k^{-}), \deputyUncertainty_i(k^{-})) \rightarrow (\hat{\deputyState}_i(k^{+}), \deputyUncertainty_i(k^{+})).$$
As a result, the size of the covariance matrices of observed deputies will necessarily decrease $f[\deputyUncertainty_i(k^{+})] \leq f[\deputyUncertainty_i(k^{-})]$ and the updated belief state will be recursively supplied to the propagator until the next available measurement update. 
Here $f[\cdot]$ is any metric quantifying the size of the covariance matrix or its representative hyperellipsoid.
However, because the sensor has a limited FOV, it may not be possible for the chief to observe every deputy at once.
Therefore, the chief must maneuver to view each deputy to maintain a level of certainty in the state estimate,
that is 
$f[\deputyUncertainty_i(k)] \leq \epsilon$
for all $k \geq 0$,
where $\epsilon$ is some positive value determined by the sensor FOV.

Then, the local catalog maintenance problem is the following.
Consider a chief satellite
with dynamic function $\dynfunc$ and constraint function $\constraintfunc$,
a sensor parametrized by pointing vector $\pointing^{\bodyframe}$ and area of view $\aov$.
Given $d$ deputies
with initial estimate $\hat{\initial} \in \stateSpace$ 
and initial covariances $\deputyUncertainty^0 \in \reals^{pd \times pd}$,
and initial chief state $\initial_0 \in \stateSpace_0$,
find a control function $\controlfunc : \stateSpace_0 \times \stateSpace \times \reals^{pd \times pd} \times \nnreals \to \inputSpace$
such that for all $k \geq 0$,
$\constraintfunc_i(\traj_{\controlfunc, \initial_0}(k), \controlfunc(\traj_{\controlfunc, \initial_0}(t))) \leq 0$ for all $i \in \{1, \dots, r\}$, and
$f[\deputyUncertainty_j(k)] \leq \epsilon$.

\paragraph{Chief attitude dynamics}
Consider a chief spacecraft in a circular orbit centered on the earth. While it does not have thrusters for translational motion, it can change its orientation using gas jet thrusters that produce external torques about the chief's center of mass. The chief's attitudinal state is given by the Euler angle set $\chiefStatex :=[\yaw,\ \pitch,\ \roll]^{\top}$, respectively denoted yaw, pitch, and roll, that parameterizes the rotation matrix from the body frame to the inertial frame $\rotmat{\bodyframe}{\inertialframe}=\rotmat{\bodyframe}{\inertialframe}^{\roll}\rotmat{\bodyframe}{\inertialframe}^{\pitch}\rotmat{\bodyframe}{\inertialframe}^{\yaw}$ using the 3-2-1 rotation sequence. The analogous angular velocities are
$\angularvel{\inertialframe}{\bodyframe}{\bodyframe} := [\omega_1, \ \omega_2, \ \omega_3]^{\top}$.
The chief inputs are given by the torques in the body frame $\chiefInput^{\bodyframe} := [\chiefInput_1, \ \chiefInput_2, \ \chiefInput_3]^{\top}$.
The rotation kinematics for the chief's body-fixed frame $\bodyframe$ is given by 
\begin{equation}
    \label{eq:rotation_velocity_b}
    \dotrotmat{\bodyframe}{\inertialframe} = \rotmat{\bodyframe}{\inertialframe}\skewMat{\angularvel{\inertialframe}{\bodyframe}{\bodyframe}}.
\end{equation}
The Euler angle set (locally) evolves according to
\begin{align}
\label{eq:euler_angles_kinematics_b}
\dot{\Gamma} &= \begin{pmatrix}
-\cos(\yaw)\tan(\pitch) & -\sin(\yaw)\tan(\pitch) & -1 \\
\sin(\yaw) & -\cos(\yaw) & 0 \\
-\cos(\yaw)\sec(\pitch) & -\sin(\yaw)\sec(\pitch) & 0 \\
\end{pmatrix} \angularvel{\mathcal{I}}{\B}{\B}.  
\end{align}

Assuming that the chief spacecraft can be modeled as a rigid body and has an inertia matrix $\chiefInertia$ as measured in the body-fixed frame, the application of Euler's second law of motion gives the attitude dynamics
\begin{equation}
    \label{eq:deputy_slew}    
    \dotangularvel{\inertialframe}{\bodyframe}{\bodyframe} = -\chiefInertiaInv\skewMat{\angularvel{\inertialframe}{\bodyframe}{\bodyframe}}[\chiefInertia\angularvel{\inertialframe}{\bodyframe}{\bodyframe}] + \chiefInertiaInv\chiefInput^{\bodyframe}.
\end{equation}
The angular velocity of the $\bodyframe$ frame relative to Hill's frame $\hillsframe$ using the property of relative angular velocity summation is 
$$\angularvel{\hillsframe}{\bodyframe}{\inertialframe} = \angularvel{\inertialframe}{\bodyframe}{\inertialframe} - \angularvel{\inertialframe}{\hillsframe}{\inertialframe}$$
with the time rate derivatives given as 
\begin{equation}
\label{eq:rel_ang_vel_inertial}
\dotangularvel{\hillsframe}{\bodyframe}{\inertialframe} = \dotangularvel{\inertialframe}{\bodyframe}{\inertialframe} - \dotangularvel{\inertialframe}{\hillsframe}{\inertialframe}.
\end{equation}
Using \eqref{eq:rotation_velocity_b} and $\angularvel{\A}{\B}{\B} = \rotmat{\A}{\B} \angularvel{\A}{\B}{\A}$ a manipulation of \eqref{eq:rel_ang_vel_inertial} yields
\begin{equation}
\label{eq:relframe_quat}
\dotangularvel{\hillsframe}{\bodyframe}{\bodyframe} = \dotangularvel{\inertialframe}{\bodyframe}{\bodyframe} - \rotmat{\hillsframe}{\bodyframe}\dotangularvel{\inertialframe}{\hillsframe}{\hillsframe} - \skewMat{\angularvel{\inertialframe}{\bodyframe}{\bodyframe}}\angularvel{\hillsframe}{\bodyframe}{\bodyframe}.
\end{equation}
Note that from the circular chief orbit assumption we have $\dotangularvel{\inertialframe}{\hillsframe}{\hillsframe} = \mathbf{0}$ and that $\angularvel{\inertialframe}{\hillsframe}{\hillsframe} = (0 \ 0 \ \meanmotion)^{\top}$ is the constant angular velocity of the Hill's frame relative to the inertial frame where
$\meanmotion$ is the constant mean motion parameter. From the assumption that the chief-fixed inertia matrix is aligned with its principal axes we have $$\chiefInertia = \begin{pmatrix}
J_1 & 0 & 0 \\ 0 & J_2 & 0 \\ 0 & 0 & J_3
\end{pmatrix}.$$
The combination of \eqref{eq:deputy_slew} and \eqref{eq:relframe_quat} yields the angular velocity dynamics as
$$\dotangularvel{\hillsframe}{\bodyframe}{\bodyframe} = -\chiefInertiaInv\skewMat{\angularvel{\inertialframe}{\bodyframe}{\bodyframe}}[\chiefInertia\angularvel{\inertialframe}{\bodyframe}{\bodyframe}] + \chiefInertiaInv\chiefInput^{\bodyframe} - \skewMat{\angularvel{\inertialframe}{\bodyframe}{\bodyframe}}\angularvel{\hillsframe}{\bodyframe}{\bodyframe}.$$
Using the relation $\angularvel{\inertialframe}{\bodyframe}{\bodyframe} = \angularvel{\hillsframe}{\bodyframe}{\bodyframe} + \angularvel{\inertialframe}{\hillsframe}{\bodyframe}$ and cross product properties the above expression becomes
\begin{align}
\label{eq:shortform_dynamics}
\begin{split}
\dotangularvel{\hillsframe}{\bodyframe}{\bodyframe} &= -\chiefInertiaInv\big(\skewMat{\angularvel{\hillsframe}{\bodyframe}{\bodyframe}}[\chiefInertia\angularvel{\hillsframe}{\bodyframe}{\bodyframe}] + \skewMat{\angularvel{\hillsframe}{\bodyframe}{\bodyframe}}[\chiefInertia\angularvel{\inertialframe}{\hillsframe}{\bodyframe}] + \\ & \qquad \qquad \qquad \qquad
\skewMat{\angularvel{\inertialframe}{\hillsframe}{\bodyframe}}[\chiefInertia\angularvel{\hillsframe}{\bodyframe}{\bodyframe}] + \skewMat{\angularvel{\inertialframe}{\hillsframe}{\bodyframe}}[\chiefInertia\angularvel{\inertialframe}{\hillsframe}{\bodyframe}]\big) + \skewMat{\angularvel{\hillsframe}{\bodyframe}{\bodyframe}}\angularvel{\inertialframe}{\hillsframe}{\bodyframe} + \chiefInertiaInv\chiefInput^{\bodyframe}.
\end{split}
\end{align}

\paragraph{Chief constraints}
The chief also has some constraints it must follow while performing the inspection. 
Typically, the torque inputs cannot exceed some $\chiefInput_{\max}$, and the angular velocity must be maintained under some $\omega_{\max}$. 
To maintain safety, the sensor cannot point in the direction of the (assumed static) sun vector $\sunvector$ given in $\inertialframe$. Recall that $\rotmat{\bodyframe}{\inertialframe}$ denotes the rotation matrix from the chief's body-fixed frame $\bodyframe$ to the inertial frame $\inertialframe$.
Then, the constraint function is given as 
\begin{equation}
    \constraintfunc(\chiefStatex, \chiefInput) = \begin{pmatrix}
    h_1(\chiefStatex,\chiefInput) \\
    h_2(\chiefStatex,\chiefInput) \\
    h_3(\chiefStatex,\chiefInput) \\
    h_4(\chiefStatex,\chiefInput) \\
    h_5(\chiefStatex,\chiefInput) \\
    h_6(\chiefStatex,\chiefInput) \\
    h_7(\chiefStatex,\chiefInput) \\
    \end{pmatrix} = \begin{pmatrix}
    \sunvector^{\top}(\rotmat{\bodyframe}{\inertialframe} \pointing_{\bodyframe}) - \cos(\aov) \\
    \|\omega_1\| - \omega_{\max} \\
    \|\omega_2\| - \omega_{\max} \\
    \|\omega_3\| - \omega_{\max} \\
    \|\chiefInput_1\| - \chiefInput_{\max} \\
    \|\chiefInput_2\| - \chiefInput_{\max} \\
    \|\chiefInput_2\| - \chiefInput_{\max} \\
    \end{pmatrix}.
\end{equation}
If $\constraintfunc_1(\chiefStatex, \chiefInput) \leq 0$,
then the angle between $\sunvector$ and $\pointing^{\inertialframe} = \rotmat{\bodyframe}{\inertialframe}\pointing^\bodyframe$ is at least $\aov$,
meaning the sun vector is not pointing in the sensor FOV.
If $\constraintfunc_i(\chiefStatex, \chiefInput) \leq 0$, $i \in \{2,3,4\}$, then $\|\omega_j\| \leq \omega_{\max}$ for $j \in \{1,2,3\}$.
If $\constraintfunc_i(\chiefStatex, \chiefInput) \leq 0$, $i \in \{5,6,7\}$, then $\|\chiefInput_j\| \leq \chiefInput_{\max}$ for $j \in \{1,2,3\}$.

\paragraph{Deputy relative motion dynamics}
The state of the deputies are described in $\hillsframe$,
and the translational dynamics of the deputies are described by the Clohessy-Wiltshire equations.
The state of a deputy $i \in \{1, \dots, d\}$ is given by $\deputyState_i = [r^i_x \ r^i_y \ r^i_z \ \dot{r}^i_x \ \dot{r}^i_y \ \dot{r}^i_z]^{\top}$, where
$r^i_x$, $r^i_y$, and $r^i_z$ are the $x$-, $y$-, $z$-positions of deputy $i$ in $\hillsframe$ respectively.
The Clohessy-Wiltshire equations can be written in state-space form as in~\eqref{eq:A-matrix}.

\begin{equation}
    \dot{\deputyState}_i = \begin{bmatrix}
        0 & 0 & 0 & 1 & 0 & 0 \\
        0 & 0 & 0 & 0 & 1 & 0 \\
        0 & 0 & 0 & 0 & 0 & 1 \\
        3 \meanmotion^2 & 0 & 0 & 0 & 2 \meanmotion & 0 \\
        0 & 0 & 0 & 2 \meanmotion & 0 & 0 \\
        0 & 0 & -\meanmotion^2 & 0 & 0 & 0
    \end{bmatrix} \deputyState_i
    \label{eq:A-matrix}
\end{equation}
Each deputy's relative state trajectory $\deputyState_i$ is uncontrolled and evolves under the following assumptions:
(i) $\dot{r}^i_y = -2 \meanmotion r^i_x$ and
(ii) $\dot{r}^i_x = \frac{1}{2} \meanmotion r^i_y$.
Condition (i) ensures that each deputy is in a periodic, closed natural motion trajectory (NMT),
and condition (ii) ensures that the NMT is a closed ellipse centered at the origin of $\hillsframe$.

\paragraph{Observation condition}
Recall from before that the deputy belief states are updated whenever an observation is made. Given a deputy state $\deputyState_i$, chief state $\chiefStatex$,
with a sensor characterized by the pointing vector in the body frame $\pointing^\bodyframe$ and the viewing angle $\aov$.
An observation occurs if $$\cos^{-1}(\rotmat{\bodyframe}{\hillsframe} \pointing^{\bodyframe} \cdot (\deputyState_i \restr \hillsframe)) \leq \aov$$ where $\rotmat{\bodyframe}{\hillsframe} = \rotmat{\inertialframe}{\hillsframe}\rotmat{\bodyframe}{\inertialframe}^{\top}.$


\section{Methodology}\label{sec: methodology}
\begin{figure}[h]
\centering
\includegraphics[width=0.75\textwidth]{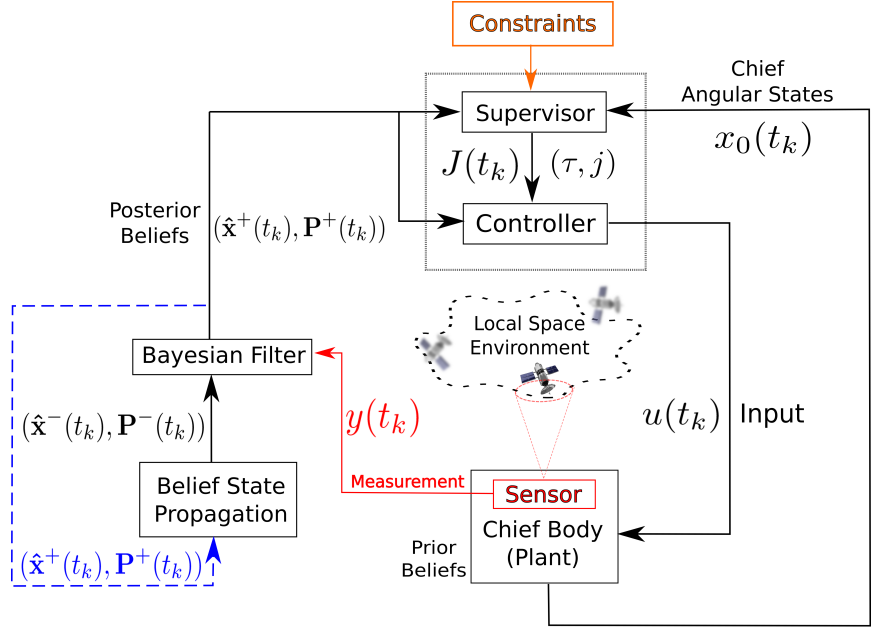}
\caption{Visual depiction of the proposed catalog maintenance operation. Note that the sensor component is embedded within the plant component. The blue dashed line indicates the recursive nature of the Bayesian filter, where updated belief states are supplied to the propagation block for the next time step.}
\label{fig:flowchart}
\end{figure}

To solve the local catalog maintenance problem, this work proposes a multi-layered closed-loop control strategy (see Fig.\ref{fig:flowchart}) that incorporates continuous dynamics with impulsive zero-order hold controls applied at discrete time instances. That is, at a given time instant $t_k$, the state estimator fuses predicted deputy belief states $(\mathbf{\hat{x}}^{-}(t_k), \mathbf{P}^{-}(t_k))$ with available observation data $y(t_k)$ to yield posterior deputy belief states $(\mathbf{\hat{x}}^{+}(t_k), \mathbf{P}^{+}(t_k))$ through a Bayesian update scheme such as the commonly-used extended Kalman filter or unscented Kalman filter.

\subsection{Supervisor Algorithm}
These updated beliefs are fed into a higher-level autonomous ``supervisor" which accounts for (i) deputy belief state information gaps (ii) prespecified constraints and (iii) the current angular states of the chief. The supervisor amalgamates this information and provides decision-making capabilities to a lower-level controller, which drives the chief orientation trajectory by administering torque inputs at discrete timesteps. 
Recall the local catalog maintenance problem where a chief, with a sensor parametrized by
$\pointing^{\bodyframe}$ and $\aov$,
must maintain state estimates of $d$ deputies.
The objective of the supervisor algorithm is to select the ``optimal" target deputy to track based on some cost function notated by $\mathcal{J}$.
A detailed breakdown of the supervisor algorithm can be seen in Algorithm \ref{alg: supervisor_algorithm}.
Given the state estimates-covariance pairs of the deputies $\{(\hat{\deputyState}_i, \deputyUncertainty_i)\}_{i=1}^{d}$
and the current deputy $j$.
Additionally, we introduce a new variable $\switchtime$ which ``remembers'' the time that the supervisor switched to deputy $j$.
The output is the deputy $j'$ to be viewed as well as $\switchtime'$.
Then, the supervisor algorithm is the following.
If some $\Delta$ time has not passed since the last switching time, or the entropy of the current deputy has not fallen below some threshold $\epsilon$,
the current deputy $j$ and current $\switchtime$ is returned.
If not, then the entropy of all the deputies is computed.
The deputy to be viewed is chosen to be the one with maximum entropy, and $\switchtime$ is updated to the current time.
The supervisor cost function in this work is considered to be the Shannon entropy of the covariance matrix, which is assumed to be Gaussian; other metrics such as measures of uncertainty relating to the deputies' belief states, such as the $\det$ or $\trace$ of the covariance matrix, or a measure of information gain post-fusion, such as the posterior Fisher information \cite{ly2017tutorial} may also be utilized.
The cost value for each deputy is calculated, and then the deputy with the highest cost value is selected to be the target deputy $j$.
However, to prevent the possibility of instantaneous switching, a form of hysteresis is applied.
A target switch is only triggered when the cost of the current target deputy falls below some threshold, the potential other target deputy crosses over the same threshold, and two target switches cannot occur within a time threshold, $\delta \geq \gamma$, of each other, where $\delta$ is the time differential between switches, and $\gamma$ is the desired time hysteresis threshold.

\begin{algorithm}
\KwInput{$\{(\hat{\deputyState}_i,\deputyUncertainty_i)\}_{i=1}^d$, 
$j$,
$\switchtime$}
\KwOutput{$j'$, $\switchtime'$}

$ \gets \textsf{CurrentTime}$ \\
\If{$(t-\switchtime \leq \Delta) \ \vee \ (\frac{k}{2}(1 + \log(2 \pi)) + \log(|\deputyUncertainty_i|) > \epsilon)$}{
$j' \gets j$ \\
$\switchtime' \gets \switchtime$}

\Else{
    \For{$i = 1, \dots, d$}{
    $\cost_i \gets \frac{k}{2}(1 + \log(2 \pi)) + \log(|\deputyUncertainty_i|)$
    }
    $j' \gets \arg \max_{i \in \{1, \dots, d\}} \cost_i$ \\
    $\switchtime' \gets t$
}

\Return $j'$, $\switchtime'$
\caption{Supervisor algorithm}
\label{alg: supervisor_algorithm}
\end{algorithm}









\subsection{Angle-Driven MPC}
We implement a state tracking version of the model predictive controller outlined in \eqref{prob:MPC}. Here we let $c(x_k, u_k) := x_k$. Let there be two positive definite square matrices $W_1 \in \reals^{2 \times 2}$ and $W_2 \in \reals^{3 \times 3}$ and an Euler angle vector $z := [\yaw, \ \pitch]^{\top}$. Note that the angle pair $\yaw$ and $\pitch$ were chosen due to the mapping from the azimuth-elevation space. The sequence over discrete steps from instance $k$ is denoted as $\textbf{z}_{k} := \{z_{0}, z_{1}, \dots, z_{i}, \dots, z_{N-1}\}$ with a reference angle trajectory (i.e. angle sequence associated with a current designated target) is denoted as $\textbf{z}^{r}_{k} := \{z^{r}_{0}, z^{r}_{1}, \dots, z^{r}_{i}, \dots, z^{r}_{N-1}\}$. 
The MPC problem is defined with a quadratic stage cost as
\begin{align}
\label{prob:MPC_method}
\begin{split}
\underset{\textbf{u}_k}{\textnormal{minimize}} & \ J(\textbf{z}_k,\textbf{u}_k) := \sum_{i=0}^{N-1}(\textbf{z}_{k}(i) - \textbf{z}^{r}_{k}(i))^{\top}W_1(\textbf{z}_{k}(i) - \textbf{z}^{r}_{k}(i)) + (\textbf{u}_{k}(i))^{\top}W_2(\textbf{u}_{k}(i))\\
\textnormal{subject to} &\  z_0 = z_k \\
                        &\  z_{i+1} = f_z(z_i,u_i), \quad i=0,\dots, N-1 \\
                        &\ (z_{i}, u_{i}) \in \mathscr{X} \times \mathscr{U}, \quad i=0,\dots, N-1. \\
\end{split}
\end{align}
Here, $f_z$ is equivalent to the dynamics expressed in \eqref{eq:euler_angles_kinematics_b} and \eqref{eq:shortform_dynamics}.




\section{Results}
\label{sec:results}

To demonstrate the efficacy of the proposed methodology, results of two example scenarios are presented - a 3 deputy case and a 10 deputy case. 
For this sample, the target deputy is selected via the Supervisor Algorithm detailed in Algorithm \ref{alg: supervisor_algorithm}.
The MPC algorithm then uses the azimuth-elevation tracks of deputy $j'$ as the reference trajectory. The CasADi optimization toolbox is used to solve the nonlinear optimal control problem \cite{Andersson2019} using a direct single shooting method.
The MPC cost function weighting matrices were set as $\mathrm{W}_{1}=\idMat{2}$ and $\mathrm{W}_{2}=\idMat{3}$.
The torque constraint was set to be $u_{max}=2\pi\ \text{Nm}$, the angular velocity constraint was set as $\omega_{max}=\pi\ \text{rad/s}$
The example simulation consists of three deputy agents following elliptical NMTs about the chief.
The chief's initial attitude and angular velocity were randomly generated with the initial angular velocity being restricted to a bounded box of $\pi/2\ \text{m/s}$ about the origin.
Because the focus of this work is on target selection and control, the initial estimates were set to the truth values and the associated covariances were set to be $\deputyUncertainty_{0} = \beta \idMat{n}$ where $\beta$ is a randomly generated parameter $10\leq\beta\leq100$.
If discrepancies in the initial truth state and state estimate were considered, the complexity of the problem increases dramatically.
Elements of target search are introduced if the discrepancy between truth position and estimated position grows sufficiently large.
This alone introduces the question: what bound on initial estimate error is feasible for estimate convergence?
This question along with related questions are too broad in scope to be considered here and have consequently been left for future work.
However, the relevancy and significance of tackling such a problem cannot be overstated.

\begin{figure}
    \centering
    \includegraphics[width=0.75\textwidth]{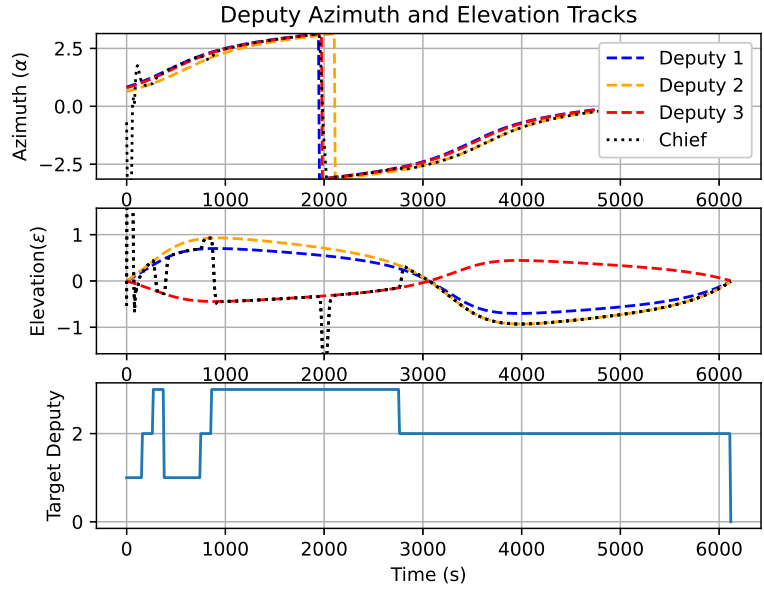}
    \caption{The MPC tracks the reference azimuth and elevation from the deputy selected by the supervisory algorithm.}
    \label{fig: extended_abstract_example}
\end{figure}

\begin{figure}
    \centering
    \includegraphics[width=0.75\textwidth]{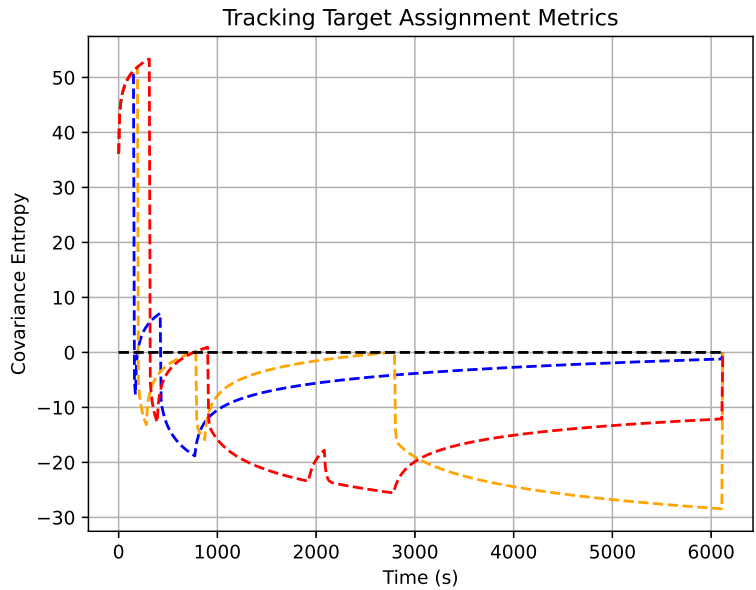}
    \caption{Covariance entropy as the chief observes each target deputy. It is shown that once each entropy value falls below the threshold, the control algorithm is able to maintain the entropy below the threshold.}
    \label{fig: entropy_tracks}
\end{figure}

\begin{figure}
    \centering
    \includegraphics[width=0.75\textwidth]{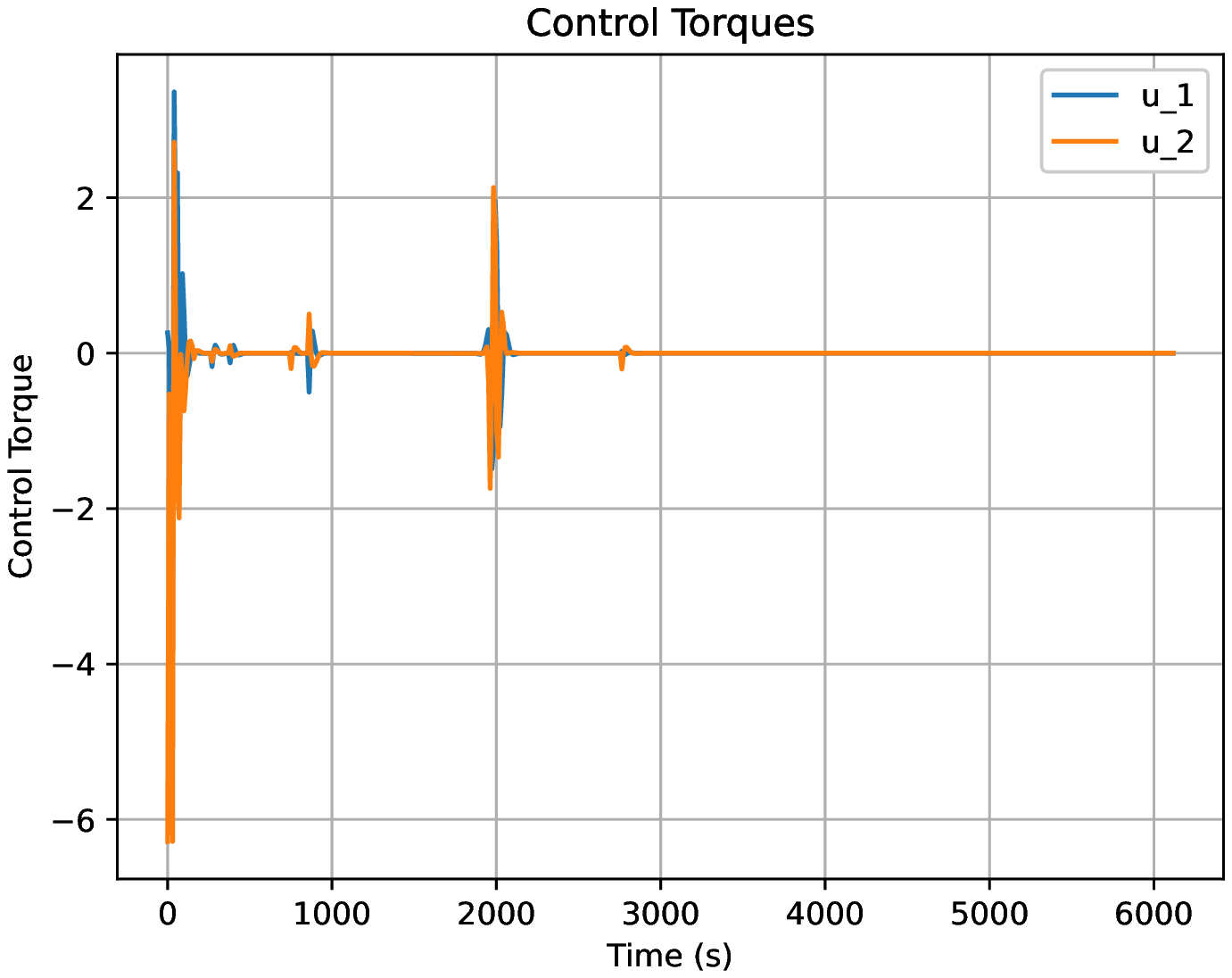}
    \caption{Applied torque for the 3 deputy case.}
    \label{fig: control_torques}
\end{figure}

Figure \ref{fig: extended_abstract_example} depicts azimuth-elevation tracks of the deputy spacecraft and the ability of the chief to track each deputy. 
The orientation of the chief switches rather frequently at the beginning of the simulation while the covariance matrices begin to settle below the specified threshold of $\epsilon=0$ on the covariance entropy seen in Figure \ref{fig: entropy_tracks}. 
To prevent instantaneous switching, a switch is only triggered when the entropy of the associated covariance matrix falls below this threshold, in and the entropy of the covariance of any other deputy exceeds the same threshold.
A time hysteresis is also applied indicating the target agent cannot switch within 100 seconds of one another such that $\Delta = 100$ from Algorithm \ref{alg: supervisor_algorithm}.
It is shown that as the simulation progresses, the chief switches less frequently because these threshold are crossed less often as the chief becomes more certain of the trajectory of each deputy.
Figure \ref{fig: control_torques} is the resulting control torque to track the reference trajectory. 
Uncharacteristically large spikes are present at about 1,750 seconds when the azimuth of the reference trajectory crosses from $\pi$ to $-\pi$, causing the chief's attitude to wrap around.
This particular cases is present due to the parameterization of the attitude, and for this reason future work is needed to adapt the presented methodology to attitude parameterizations that do not present this discontinuity such as rotation matrices and unit quaternions.

\begin{figure}
    \centering
    \includegraphics[width=0.75\textwidth]{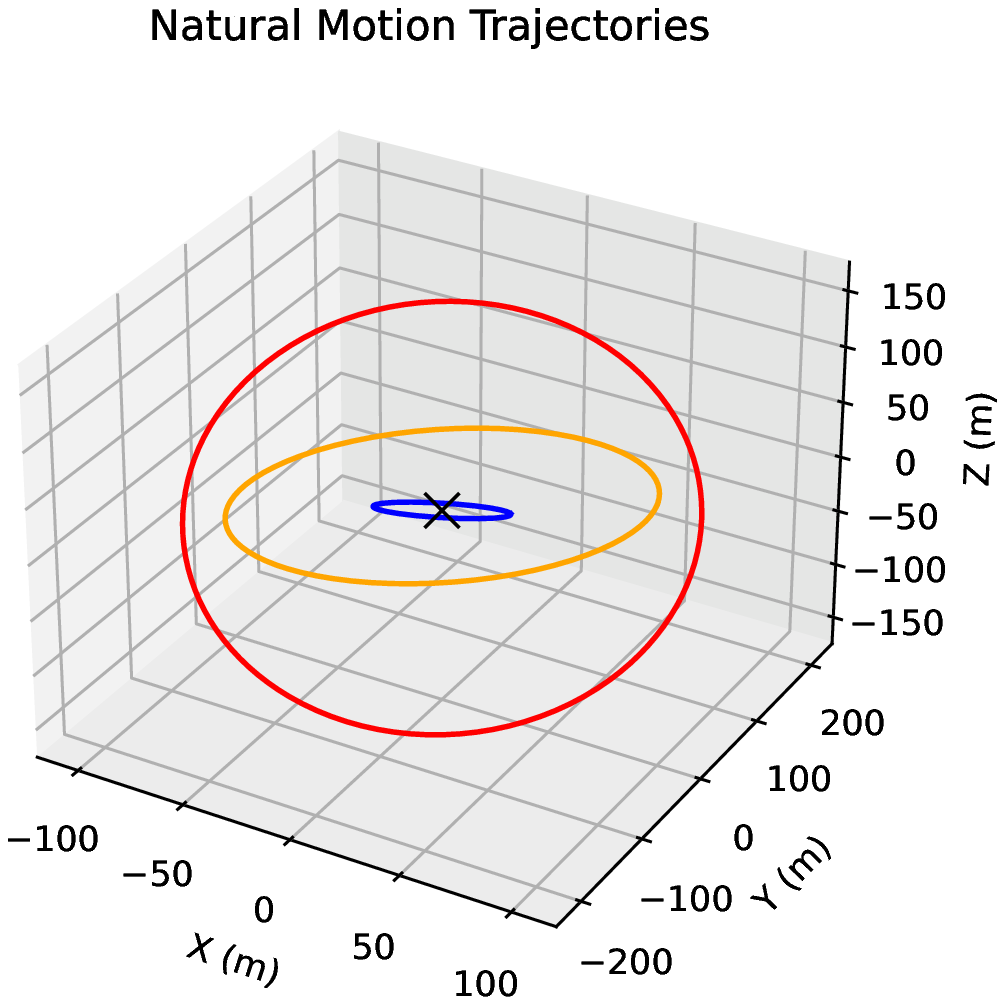}
    \caption{NMTs that generate the azimuth-elevation tracks.}
    \label{fig: generating_nmt}
\end{figure}

\begin{figure}
    \centering
    \includegraphics[width=0.75\textwidth]{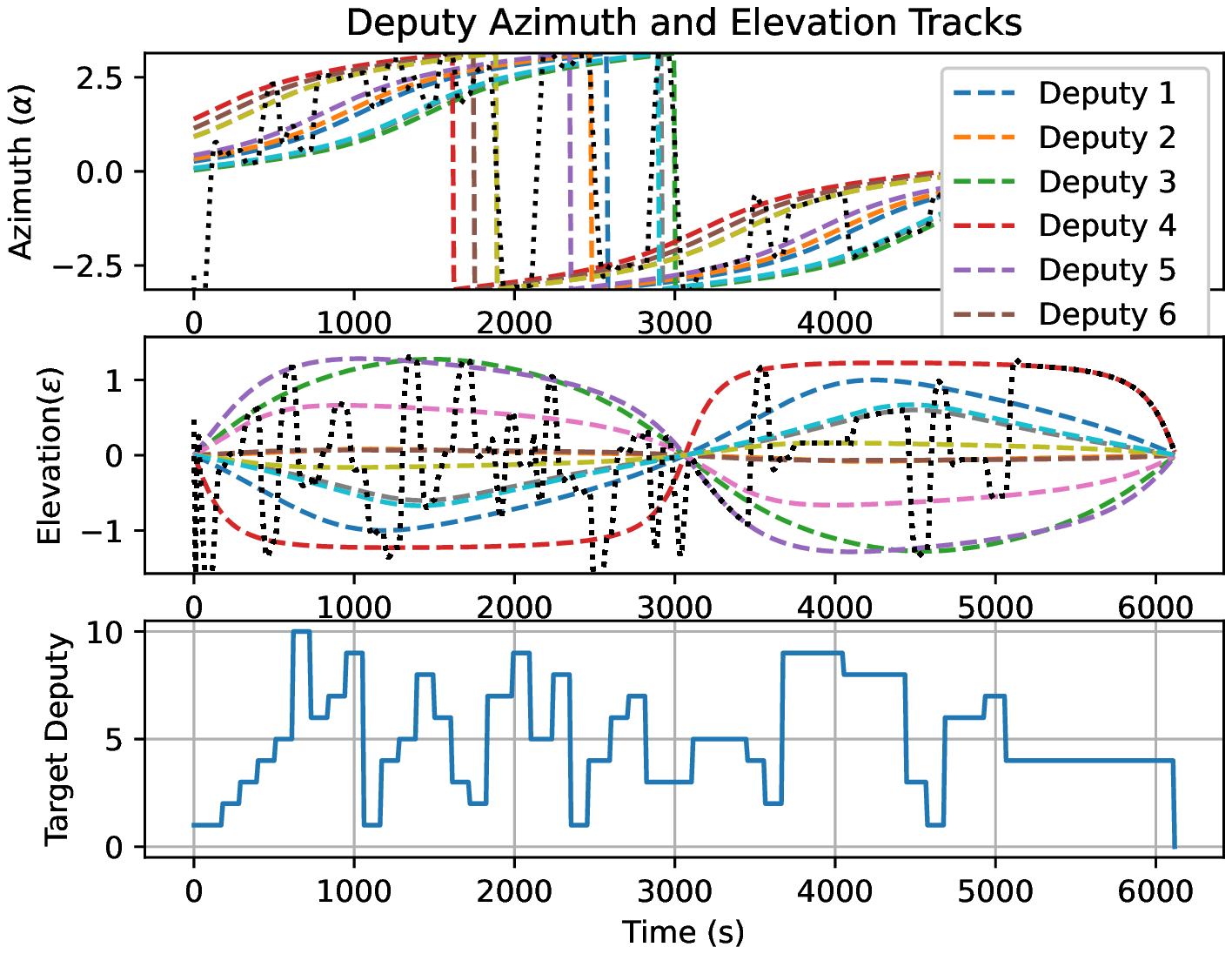}
    \caption{The algorithms are capable of tracking the reference azimuth and elevation track when more deputies require tracking.}
    \label{fig: ae_tracks_10}
\end{figure}

\begin{figure}
    \centering
    \includegraphics[width=0.75\textwidth]{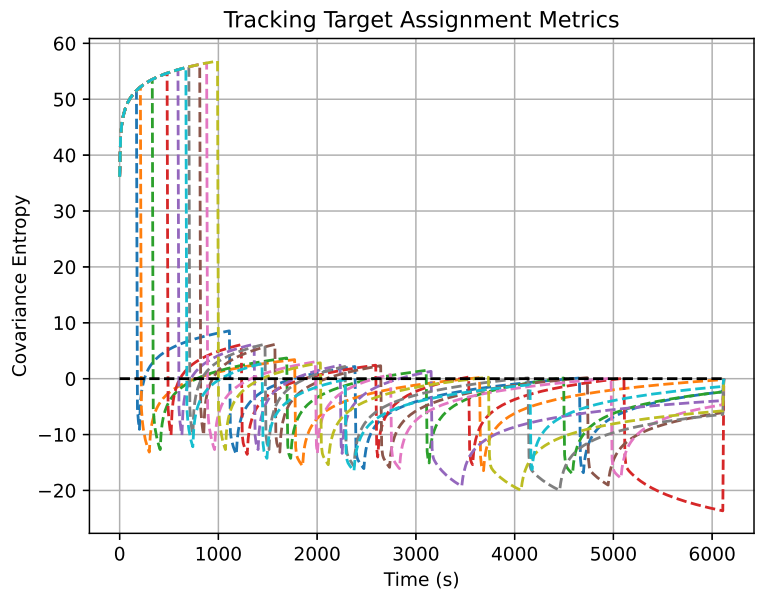}
    \caption{The algorithms are also able to maintain the covariance entropy below the desired threshold, even though it takes longer to achieve this level of estimator confidence.}
    \label{fig: entropy_tracks_10}
\end{figure}

\begin{figure}
    \centering
    \includegraphics[width=0.75\textwidth]{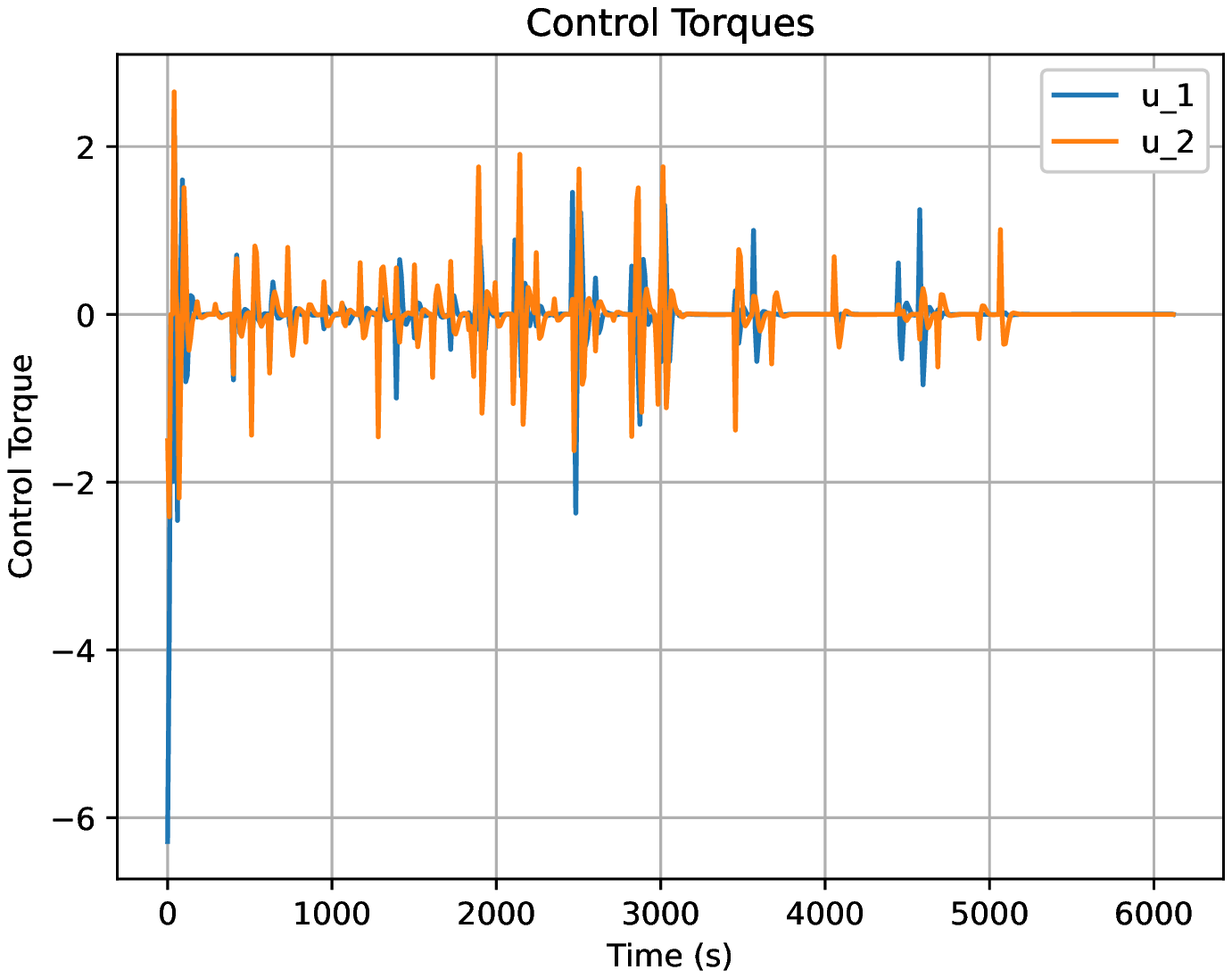}
    \caption{Applied torque for the 10 deputy case.}
    \label{fig: control_torques_10}
\end{figure}

The performance of the algorithm was also tested on a 10 deputy case, shown in Figure \ref{fig: ae_tracks_10}, Figure \ref{fig: entropy_tracks_10}, Figure \ref{fig: control_torques_10}.
In Figure \ref{fig: ae_tracks_10} the MPC is able to correctly track the reference trajectory specified by the supervisor algorithm even with increased deputy count.
However, because of the increased number of deputies, the chief is not able to linger on each deputy for as long and must quickly move on to track the next deputy. 
Despite the increased workload, the chief is still able to achieve the desired level of confidence in each state estimate indicated by Figure \ref{fig: entropy_tracks_10}.
But, each entropy stays closer to the threshold when compared to the 3 deputy case.
Furthermore, the chief must apply a greater amount of torque throughout the duration of the the mission seen in Figure \ref{fig: control_torques_10}.
This is a direct result of the supervisor algorithm greedily selecting the next reference deputy based on the information entropy. 
It is possible to include control effort in the supervisor algorithm to help mitigate the effort applied by the chief.

\section{Conclusions}
This paper has provided a solution to the local catalog maintenance problem, which involves a controllable chief spacecraft performing state estimation across multiple deputy agents. The maintenance problem is, at its core, a decision-making problem. Thus, this work proposes a hierarchical supervising algorithm which autonomously decides which of the nearby deputies must be observed. A model predictive controller is implemented to generate a trajectory of torque inputs that are optimal with respect to minimizing the deputies' state uncertainties. We demonstrate through numerical simulation that this supervisor scheme succeeds in handling both short-term and long-term estimation requirements.

\newpage

\bibliographystyle{AAS_publication}
\bibliography{fullpaper}

\end{document}